\documentclass{JHEP}
\usepackage{epsfig}
\usepackage[active]{srcltx}

\newcommand{\be}{\begin{equation}}
\newcommand{\ee}{\end{equation}}
\newcommand{\bea}{\begin{eqnarray}}
\newcommand{\eea}{\end{eqnarray}}
\newcommand{\bean}{\begin{eqnarray*}}
\newcommand{\eean}{\end{eqnarray*}}

\def\abs#1{\left| #1\right|}
\def\beq{\begin{equation}}

\def\bra#1{\left\langle #1\right|}
\def\eeq{\end{equation}}

\def\ket#1{\left| #1\right\rangle}
\def\Re{\mathop{\rm Re}}

\def\Arcsech{{\rm arcsech}}
\def\Arctanh{{\rm arctanh}}
\def\csch{{\rm csch}}

\def\eref#1{(\ref{#1})}
\def\bit#1{\begin{array}{c} \\ \\ \end{array} \hspace{#1 cm}}

\preprint{MIT-CTP-3243\\ {\tt hep-th/0202176}}

\title{The Spectrum of the Neumann Matrix with Zero Modes}

\author{Bo Feng, Yang-Hui He and Nicolas Moeller
\\
Center for Theoretical Physics,
\\Massachusetts Institute of Technology,\\ Cambridge, MA 02139, USA\\
\email{fengb,yhe,moeller@ctp.mit.edu}
}

\abstract{We calculate the spectrum of the matrix $M^{\prime}$ of 
Neumann coefficients of the Witten vertex, expressed in the oscillator 
basis including the zero-mode $a_0$. We find that in addition to the
known continuous spectrum inside $[-\frac13,0)$
of the matrix $M$ without the zero-modes, 
there is also an additional eigenvalue inside $(0,1)$. 
For every eigenvalue, there is a
pair of eigenvectors, a twist-even and a twist-odd. We give
analytically these 
eigenvectors as well as the generating function for their components.
Also, we have found an interesting critical parameter $b_0 = 8 \ln
2$ on which the forms of the eigenvectors depend.
}

\begin{document}
\section{Introduction}

Vacuum String Field Theory (VSFT) was proposed by Rastelli, Sen and 
Zwiebach in \cite{0012251} - \cite{0111153} 
as the expansion of Witten's Open String Field Theory \cite{Witten_boson} 
around the 
non-perturbative vacuum. They conjectured that the kinetic operator of VSFT 
is pure ghost after a suitable (possibly singular) field redefinition. 
A strong support of this 
conjecture was that they could reproduce numerically, with convincing
precision,  
the correct D-brane descent relations \cite{0102112}. These descent 
relations were further 
established in the context of Boundary Conformal Field Theory 
\cite{0105168}. However, until recently, a direct algebraic derivation 
based on the properties of the Neumann coefficients has been elusive;
and so have been
the proofs of other conjectures, such as the equality between the algebraic 
and the geometric sliver, or the form of the pure-ghost kinetic operator 
around the stable vacuum.

These proofs all came up very recently, shortly 
after Rastelli, Sen and Zwiebach solved the spectrum of the matrix $M$
of Neumann
coefficients \cite{Spec}. They found that the spectrum is continuous in the 
range $[-{1 \over 3}, 0)$; every eigenvalue in this interval is doubly
degenerate,  
except for $-{1 \over 3}$ which is single and {\em twist-odd}. 
They gave a complete solution by 
finding the density of eigenvalues and the expressions of the corresponding 
eigenvectors. This result turned out to be a key tool for 
doing exact calculations in VSFT. Indeed, using the form of 
the spectrum of $M$, Okuyama 
\cite{Spec_Oku_1} proved that 
the ghost kinetic operator of VSFT is given by the ghost field 
$c$ evaluated at 
the string midpoint, as was already expected \cite{Hata:2001sq, 0111129}. 
Then in another paper \cite{Spec_Oku_2}, Okuyama also 
gave an algebraic proof that the D-brane descent relation is correctly 
reproduced. The ration of the tension of a D$p$-brane to the tension of a 
D$(p+1)$-brane can be expressed in terms of determinants of matrices
of Neumann coefficients  
\beq
R = {T_p \over 2 \pi \sqrt{\alpha^{\prime}} T_{p+1}} = 
{3 \left(V_{00}^{rr} + {b \over 2} \right)^2 \over \sqrt{2 \pi b^3}} 
{\det (1-M^{\prime})^{3 \over 4} (1+3 M^{\prime})^{1 \over 4} \over 
\det (1-M)^{3 \over 4} (1+3 M)^{1 \over 4}} \,,
\eeq
where $M$ is the matrix formed by the Neumann coefficients of the vertex 
in the oscillator basis with zero momentum, whereas $M^{\prime}$ is made out 
of the Neumann coefficients of the vertex expressed in the oscillator basis 
including the zero-mode oscillator $a_0$. The parametre $b$ is an
arbitrary constant in the definition $a_0^{\mu} := \frac12 \sqrt{b}
p^{\mu} - \frac{1}{\sqrt{b}} i x^{\mu}$ \cite{0102112}.
Although it seems, at a first look, 
that one needs the spectrum of both $M$ and $M^{\prime}$ to calculate $R$, 
Okuyama \cite{Spec_Oku_2} found an elegant way of calculating this
ratio knowing only 
the spectrum of $M$. At last, Okuda \cite{Okuda_1} proved the equality of 
the geometric sliver and the algebraic sliver 
\cite{0102112, 0106010, 0006240, KP, Furuuchi, MT}.

Because the spectrum of $M$ is such an important piece of data, it is
reasonable to  
expect that knowing the spectrum of $M^{\prime}$ will be very useful
as well.
In this paper we thus solve the problem of finding all 
eigenvalues and eigenvectors of $M^{\prime}$.

We summarize our results here: We find
that the eigenvalues of $M^{\prime}$ are given by two types, a {\bf
continuous} and a {\bf discrete} spectrum. 
The continuous eigenvalues are the same as that of $M$ and are
located in the range $[-1/3,0)$. 
The discrete eigenvalue is located in the range $(0,1)$ and is 
determined by (\ref{implicit}) (or (\ref{a_oo_det})) implicitly. 
The corresponding eigenvectors are as follows.
For every eigenvalue $\lambda \in [-1/3,0)$, we have two degenerate
eigenvectors which can be written as a twist-even (\ref{v_even}) and a
twist-odd (\ref{v_odd}).
Note that this degeneracy includes the point $\lambda = -1/3$.
For the discrete eigenvalue $\lambda \in (0,1)$, we have again two
degenerate eigenvectors, a twist-even
(\ref{iso_even}) and a twist-odd (\ref{iso_odd}). They do not have
corresponding analogues in $M$ and consist only of certain vectors
$\ket{v_e}$ and $\ket{v_o}$ defined in \eref{Spec_Okuyama_2_3.11}. 

Interestingly, we have found a critical value
$b_0=8\ln 2\approx 5.54518$ where the forms of the eigenvectors differ
slightly for $b\geq b_0$ and $b<b_0$. 
When $b<b_0$, the eigenstates for the continuous spectrum 
((\ref{v_even}) and (\ref{v_odd}))
can be considered as deformations of those
of $M$ by $\ket{v_e}$ and $\ket{v_o}$.
When $b\geq b_0$, all eigenvectors are as above
except at one point $\lambda_0
\in [-1/3,0)$ determined by (\ref{implicit}) (or (\ref{a_oo_det})).
At this particular point, the corresponding eigenvector will 
have the form given by (\ref{iso_even}) and (\ref{iso_odd})
instead of the ones given by (\ref{v_even}) and (\ref{v_odd})
for the aforementioned continuous spectrum.

This paper is organized as follows. Section 2 summarizes some of the
known properties of the matrices $M$ ad $M'$ which are key to our
derivations. Then, after a review of the method of diagonalising $M$
in Section 3, we reduce the central problem of diagonalising $M'$ into
a linear system of equations in Section 4, wherein we also present the
continuous spectrum. In Section 5, we discuss 
the analytic evaluations and behaviour of zeros of the determinant of
the linear system. Sections 6 is the highlight
of the paper where we carefully analyse the discrete spectrum of
$M'$. In Section 7 we evaluate the
so-called generating functions explicitly to obtain the components of
the eigen-vectors. Finally in Section 8, we apply our methods to
analyse the spectra of the other $M'^{rs}$ matrices.
We end with conclusions and prospects in Section 9.

\section{Notations and Some Known Results}
In this section, we recall some known results and fix the notation we
shall  
use throughout the paper. 
All relevant results can be found in \cite{Spec,Spec_Oku_1,Spec_Oku_2}. 
We emphasize here that we take $\alpha' = 1$.

\subsection{Properties of the Matrix $M$}
We first recall the definition of the matrix $M$, defined as a product
of the {\em twist matrix} $C_{mn}$ and the Neumann Coefficients
$V^{11}_{mn}$ for the star product in open bosonic string field
theory:
\[
\left( M \right)_{mn} := \left( CV^{11} \right)_{mn}; \qquad C_{mn} :=
	(-1)^m \delta_{mn}.
\]
In \cite{Spec}, it was found that the eigenvectors of $M$ can be
written as 
\be
\label{k-def} 
\ket{k}=(v_1^{k},v_2^{k},v_3^{k},...)^T,
\ee
with eigenvalue
\be
\label{M_eig} M(k)=-\frac{1}{1+2\cosh \frac{\pi k}{2}}.
\ee
The components $v_i^{k}$ can be found from the generating function
\be
\label{f_gen} 
f_k(z)=\sum_{n=1}^{+\infty} \frac{v_n^{k}}{\sqrt{n}} z^n
=\frac{1}{k}(1-e^{-k\arctan z}).
\ee
We can simplify notations by defining the inner product \cite{Spec_Oku_1}
$$
\langle z\ket{k}\equiv \sum_{n=1}^{+\infty} z^n v_n^{k},
$$
where $\ket{z}\equiv (z,z^2,z^3,....)^T$ and $\bra{z}=\ket{z}^T$ is
the transpose 
of $\ket{z}$ (not hermitian conjugate). Then the generating function becomes
\be
\label{f_gen_1}
f_k(z)=\bra{z}E^{-1}\ket{k}=\langle k|E^{-1}\ket{z}
\ee 
where $E_{nm}=\sqrt{n} \delta_{nm}$. Under the twist action of $C$
defined above, we have
\be
\label{Twist}
C\ket{z}=\ket{-z}, \qquad C\ket{k}=-\ket{-k}.
\ee

The eigenvector $\ket{k}$ has very good properties, most notably the
orthogonality under the inner product\cite{Spec_Oku_2}:
\be
\label{inner} 
\langle k | p \rangle= {\cal N}(k)\delta(k-p), \qquad 
{\cal N}(k) := \frac{2}{k} \sinh(\frac{\pi k}{2}).
\ee
Using this result, we see that $\ket{k}$ forms a complete basis and 
\be
\label{identity}
{\bf 1}=\int_{-\infty}^{+\infty} dk \frac{\ket{k}\bra{k}}{{\cal N}(k)}
\ee

\subsection{The Matrix of our Concern: $M'$}
The matrix we try to diagonalize is \cite{0102112,Spec_Oku_2}
\be
\label{Spec_Okuyama_2_3.1} 
M'=\left( \begin{array}{cc} M'_{00} & M'_{0m} \\
  M'_{n0} & M'_{nm} \end{array} \right)=\left( \begin{array}{ll}
  1-\frac{2}{3} \frac{b}{ \beta} & 
  -\frac{2}{3} \frac{\sqrt{2b}}{ \beta} \bra{v_e}~~~~~~~ \\
  -\frac{2}{3} \frac{\sqrt{2b}}{ \beta} \ket{v_e} \qquad &
  M+\frac{4}{3} \frac{(-\ket{v_e}\bra{v_e}+\ket{v_o}\bra{v_0})}{ \beta} 
 \end{array} \right),
\ee
where we have defined
\bean
\beta=V_{00}^{rr}+\frac{b}{2} = \ln \frac{27}{16} + \frac{b}{2},&  & \\
\ket{v_e}=E^{-1}\ket{A_e},& &\quad \ket{v_o}=E^{-1}\ket{A_o},\\
(A_e)_n = \frac{1+(-)^n}{2} A_n, & &(A_o)_n = \frac{1-(-)^n}{2} A_n,
\eean
and $A_n$ is defined as the coefficients of the series expansion
\be
\label{A_n}
\left(\frac{1+ix}{1-ix}\right)^{1/3}  =  \sum_{n=even} A_n x^n+ i\sum_{n=odd}A_n x^n.
\ee

There are a few results concerning the states $\ket{v_e}$ and
$\ket{v_o}$ which we will use later.
We quote them from \cite{Spec_Oku_2} as
\be
\label{Spec_Okuyama_2_3.11} 
\bra{k}v_e\rangle=\frac{1}{k} \frac{\cosh(\frac{\pi k}{2})-1}
  {2\cosh(\frac{\pi k}{2})+1},~~~~\bra{k}v_o \rangle=\frac{\sqrt{3}}{k} 
\frac{\sinh(\frac{\pi k}{2})}  {2\cosh(\frac{\pi k}{2})+1}, 
\ee
and\footnote{As a byproduct of our analysis, we will actually 
	prove this identity and another one
	$\bra{v_o}\frac{1}{1-M}\ket{v_o}=\frac{3}{4} V_{oo}^{rr}$ later.} 
\be
\label{Spec_Okuyama_2_3.13} 
\bra{v_e}\frac{1}{1+3M}\ket{v_e}=\frac{1}{4}V_{00}^{rr}=
\frac{1}{4}\ln \frac{27}{16} 
\ee
The twist operation on these states are easily seen to be
\be
\label{eo_parity} C\ket{v_e}=\ket{v_e}, \qquad C\ket{v_o}=-\ket{v_o}
\ee
\section{One Simple Example}
In this section, we will use one simple example to demonstrate our
method to diagonalize the matrix $M'$ in \eref{Spec_Okuyama_2_3.1}. 
We shall use the technique in \cite{Spec_Oku_1,Spec_Oku_2}
to find the eigenvector $v$ and eigenvalue $\lambda$ of the matrix
$M$:
$$
M\cdot v= \lambda v.
$$
Using \eref{identity} we can expand $v$ into the $\ket{k}$  basis as
\be
v=\int_{-\infty}^{+\infty} dk h(k) \ket{k}.
\ee
Now we have
\bean
M\cdot v & = & M \int_{-\infty}^{+\infty} dk h(k) \ket{k}  \\
   & = & \int_{-\infty}^{+\infty} dk h(k) M(k)\ket{k}  \\
   & = &  \int_{-\infty}^{+\infty} dk \lambda h(k) \ket{k}, \\
\Rightarrow  0 & = &   \int_{-\infty}^{+\infty} dk
h(k)\left( \lambda-M(k) \right) \ket{k}.
\eean
Since the different $\ket{k}$ are independent of each other, a
na\"{\i}ve solution is
$$
h(k)\left( \lambda-M(k) \right) = 0, \qquad \forall k,
$$
giving the trivial solution $h(k)=0$.
However, we can find a non-trivial solution as follows. 
Recalling that for an arbitrary
function $f(k)$ with a zero at $k_0$ so that $f(k_0)=0$, we have
\be
\int_{-\infty}^{+\infty} dk \delta(k-k_0) f(k)=0,
\ee
we should require\footnote{In fact, it seems that equation (\ref{3.3}) does
not make sense because the
right hand side of (\ref{3.3}) is zero. However, the meaning of
(\ref{3.3}) should 
be understood as that the left hand side should have the form of right
hand side. It is in this sense that we write down this formula and use it
to solve $h(k)$. In other
words, the equation $z f(z)=0$ has solution $f(z)=a\delta(z)$ where $a$ is
an overall constant. Therefore (\ref{3.3}) can be solved as
$h(k)=a\delta(\lambda-M(k))=a' \delta(k-k_0)$. We want to thank D. Belov
for pointing out this subtle point.}

\be
\label{3.3}
h(k)\left( \lambda-M(k) \right) = \delta(k-k_0) f(k), \qquad \forall k.
\ee 
This means that we can choose
\be
h(k)=\delta(k-k_0),\quad \mbox{and} \quad \lambda-M(k)=f(k).
\ee
Therefore we can solve (recall that $f(k_0)=0$)
\be
\lambda=M(k_0)=-\frac{1}{1+2\cosh \frac{\pi k_0}{2}}
\ee
and
\be
v=\int_{-\infty}^{+\infty} dk h(k) \ket{k}=\int_{-\infty}^{+\infty} dk
\delta(k-k_0)\ket{k}=\ket{k_0},
\ee
which are the known eigenvalue and eigenvector respectively.
\section{Diagonalising $M'$: Setup and Continuous Spectrum}
After the preparation above, we can start to diagonalize the matrix
$M'$ in \eref{Spec_Okuyama_2_3.1}.
First we expand the eigenstate as 
\be
\label{vector_form} 
v=\left[ \begin{array}{c} g \\ 
  \int_{-\infty}^{+\infty} dk h(k) \ket{k}  \end{array} \right],
\ee
where $g$ is a number corresponding to the zero mode and $h(k)$ is the
coefficient of expansion on the $\ket{k}$-basis.
Then $M' \cdot v= \lambda v$ transforms into two equations
\bea
\label{v1} \lambda g & = & (1-\frac{2}{3} \frac{b}{ \beta}) g
-\frac{2}{3} \frac{\sqrt{2b}}{ \beta} 
\int_{-\infty}^{+\infty} dk h(k)\bra{v_e} k \rangle, \\
\label{v2}
\int_{-\infty}^{+\infty} dk \lambda h(k) \ket{k} & = & 
-\frac{2}{3} \frac{\sqrt{2b}}{ \beta} \ket{v_e} g+
\int_{-\infty}^{+\infty} dk h(k) M(k)\ket{k}    \\
& & +\frac{4}{3} \frac{1}{ \beta}\left( -\ket{v_e} 
  \int_{-\infty}^{+\infty} dk h(k)\bra{v_e} k \rangle+\ket{v_o}
 \int_{-\infty}^{+\infty} dk h(k) \bra{v_o} k \rangle \right). \nonumber
\eea
For later convenience, we define
\be
\label{C_eo}
{\cal C}_e[h(k)] =\int_{-\infty}^{+\infty} dk h(k)\bra{v_e} k\rangle, \qquad
{\cal C}_o[h(k)] =\int_{-\infty}^{+\infty} dk h(k)\langle v_o| k\rangle,
\ee
and solve $g$ from \eref{v1} as
\be
\label{g}
g=\frac{2\sqrt{2b}}{3\beta (1-\lambda)-2b} {\cal C}_e.
\ee

Putting \eref{g} into \eref{v2} and simplifying we obtain
\be
\label{v2_1}
\int_{-\infty}^{+\infty} dk \lambda h(k) \ket{k} =
   \int_{-\infty}^{+\infty} dk h(k) M(k)\ket{k} + 
   \frac{4(\lambda-1)}{3\beta (1-\lambda)-2b}\ket{v_e} {\cal C}_e
  +\frac{4}{3\beta}\ket{v_o} {\cal C}_o.
\ee
Now we expand $\ket{v_e},\ket{v_o}$ as
\be
\ket{v_e}=\int_{-\infty}^{+\infty} dk \ket{k}\frac{\bra{k}v_e
   \rangle}{{\cal N}(k)},\qquad 
\ket{v_o}=\int_{-\infty}^{+\infty} dk
   \ket{k}\frac{\bra{k}v_o\rangle}{{\cal N}(k)},
\ee
and get
\be
\label{v2_2} 
0 = \int_{-\infty}^{+\infty} dk \ket{k}\left( -\lambda h(k)
  + h(k) M(k)+\frac{4(\lambda-1)}{3\beta (1-\lambda)-2b} 
  \frac{\bra{k}v_e \rangle}{{\cal N}(k)}{\cal C}_e
  + \frac{4}{3\beta}\frac{\bra{k}v_o\rangle }{{\cal N}(k)} {\cal
  C}_o \right).
\ee

From the experience we gained in the previous section we should
require that
\bea
\label{v2_3}
& & \left(-\lambda + M(k) \right) h(k)
+\frac{4(\lambda-1)}{3\beta (1-\lambda)-2b} 
  \frac{\bra{k}v_e\rangle}{{\cal N}(k)}{\cal C}_e 
   + \frac{4}{3\beta}\frac{\bra{k}v_o\rangle}{{\cal N}(k)} {\cal C}_o 
  \nonumber \\ 
& = & -\delta(k-k_0)  r(k),
\eea
where $r(k)$ is an arbitrary integrable function with a zero at
$k_0$. Here we want to emphasize that at this point $k_0$ is a yet
to be determined parameter and $r(k)$, a to be determined function. We
will show later how to determine these.

Now equation \eref{v2_3} is an Fredholm integral equation of the first
kind in $h(k)$. To solve it we need to write it into the
standard form as\footnote{The term 
${ 1\over \lambda-M(k)}$ is not very well defined when
we write it in this form. However, the only physically meaningful 
quantity is the expression $\int dk h(k) \ket{k}$. When we perform the
integration, we should
choose the principal-value integration. 
This fixes the definition. We want
to thank Dmitri Belov for discussing with us about this point.}
\be
\label{v2_4} 
h(k)  = \frac{4(\lambda-1)}{3\beta (1-\lambda)-2b} 
  \frac{\bra{k}v_e\rangle}{{\cal N}(k) (\lambda-M(k))}{\cal C}_e 
   + \frac{4}{3\beta}\frac{\bra{k}v_o\rangle}{{\cal N}(k)(\lambda-M(k))}
 {\cal C}_o + \frac{\delta(k-k_0)  r(k)}{\lambda-M(k)}.
\ee

Applying the operation $\int_{-\infty}^{+\infty} dk \langle v_e\ket{k}$
on both sides of \eref{v2_4} we obtain
\be
\label{v2_5} {\cal C}_e = \frac{4(\lambda-1)}{3\beta (1-\lambda)-2b} 
  A_{ee} {\cal C}_e +\frac{4}{3\beta} A_{eo} {\cal C}_o
  + B_{e},
\ee
where we have defined
\bea
 A_{ee}(\lambda) &=&  \int_{-\infty}^{+\infty} dk
  \frac{\bra{k}v_e\rangle \langle v_e\ket{k}}{{\cal N}(k)
  (\lambda-M(k))} = \bra{v_e} \frac{1}{\lambda - M} \ket{v_e}, \\
 A_{eo}(\lambda) &=&  \int_{-\infty}^{+\infty} dk
  \frac{\bra{k}v_o\rangle \langle v_e\ket{k}}{{\cal N}(k)
  (\lambda-M(k))}  = \bra{v_e} \frac{1}{\lambda - M} \ket{v_o},  \\
 A_{oo}(\lambda) &=&  \int_{-\infty}^{+\infty} dk
  \frac{\bra{k}v_o\rangle \langle v_o\ket{k}}{{\cal N}(k)
  (\lambda-M(k))} = \bra{v_o} \frac{1}{\lambda - M} \ket{v_o} ,\\
 B_{e}(\lambda) &=&  \int_{-\infty}^{+\infty} dk 
   \frac{\delta(k-k_0)  r(k)\langle v_e\ket{k}}{\lambda-M(k)}, \\
 B_{o}(\lambda) &=&  \int_{-\infty}^{+\infty} dk 
   \frac{\delta(k-k_0)  r(k)\langle v_o\ket{k}}{\lambda-M(k)}.
\label{ab}
\eea
The integrals for $B_{e}(\lambda)$ and $B_{o}(\lambda)$ are subtly
dependent on the parametres $r(k)$ and $k_0$ and will be addressed in
Subsection 4.1. The $A$ integrals will be the subject of Section 5.

Similarly, applying $\int_{-\infty}^{+\infty} dk \langle v_o\ket{k}$ on
both sides of \eref{v2_4} we get
\be
\label{v2_6} {\cal C}_o=\frac{4(\lambda-1)}{3\beta (1-\lambda)-2b}
  A_{eo} {\cal C}_e +\frac{4}{3\beta} A_{oo} {\cal C}_o
  + B_{o}.
\ee

Equations \eref{v2_5} and \eref{v2_6} can be written in matrix form as
\be
\label{v2_7} \left[ \begin{array}{ll} 
  1-\frac{4(\lambda-1)}{3\beta (1-\lambda)-2b}A_{ee}~~~  & 
  -\frac{4}{3\beta} A_{eo}  \\
  -\frac{4(\lambda-1)}{3\beta (1-\lambda)-2b} A_{eo} &
  1-\frac{4}{3\beta} A_{oo} \end{array} \right]
 \left[ \begin{array}{l} {\cal C}_e \\ {\cal C}_o \end{array} \right]
=\left[ \begin{array}{l}  B_{e}  \\  B_{o} \end{array} \right].
\ee

Using the expression \eref{Spec_Okuyama_2_3.11} it is easy to show
(due to the odd parity of the integrand) that
$A_{eo}=0$. Therefore \eref{v2_7} is actually diagonal
\be
\label{v2_8} \left[ \begin{array}{ll} 
  1-\frac{4(\lambda-1)}{3\beta (1-\lambda)-2b}A_{ee}~~~  & 0 \\
  0 &
  1-\frac{4}{3\beta} A_{oo} \end{array} \right]
 \left[ \begin{array}{l} {\cal C}_e \\ {\cal C}_o \end{array} \right]
=\left[ \begin{array}{l}  B_{e}  \\  B_{o} \end{array} \right].
\ee
We have reduced the eigenproblem for $M'$ to the linear
system \eref{v2_8}. As we will show immediately, in obtaining nonzero
solutions for \eref{v2_8}, we determine the eigenvalue $\lambda$,
which will then fix $k_0$ and $r(k)$ accordingly.
After this,
we substitute the solutions for ${\cal C}_e, {\cal C}_o$
into \eref{v2_4},\eref{g} to give $h(k),g$, which
henceforth determines the eigenvectors by \eref{vector_form}.

Of crucial
importance is therefore the determinant of the left hand side of
\eref{v2_8},
\[
Det := \left| \begin{array}{ll} 
  1-\frac{4(\lambda-1)}{3\beta (1-\lambda)-2b}A_{ee}~~~  & 0 \\
  0 &
  1-\frac{4}{3\beta} A_{oo} \end{array} \right| = 
\left(1-\frac{4(\lambda-1)}{3\beta (1-\lambda)-2b}A_{ee}  \right) 
\left( 1-\frac{4}{3\beta} A_{oo} \right) .
\]
When $Det \ne 0$ we can have a continuous spectrum of solutions which
we address in the following. When $Det=0$, there are a
finite number of solutions which
will be the subject of Section 6.
\subsection{The Continuous Spectrum}
For the $\lambda$ values which do not make $Det$ zero, we can solve
\eref{v2_8} as
\bea
\label{Ce_1}
{\cal C}_e & = & \frac{ B_{e}}{1-\frac{4(\lambda-1)}{3\beta
(1-\lambda)-2b}A_{ee}}\equiv\frac{ B_{e}}{M_{ee}} , \\ 
\label{Co_1}
{\cal C}_o & = & \frac{ B_{o}}{1-\frac{4}{3\beta} A_{oo}}\equiv 
\frac{ B_{o}}{M_{oo}}.
\eea
We claim that only when $\lambda\in [-1/3,0)$ we can 
get a nonzero solution. The reason is as follow.
From the explicit forms of $B_{e}$ and $B_{o}$
\bea
\label{B_e_1} B_e & = &  \int_{-\infty}^{+\infty} dk \frac{1}{k} 
\frac{\cosh(\frac{\pi k}{2})-1}{2\cosh(\frac{\pi k}{2})+1}
  \frac{\delta(k-k_0)  r(k)}{\lambda-M(k)}, \\
\label{B_o_1} B_o & = &  \int_{-\infty}^{+\infty} dk\frac{\sqrt{3}}{k} 
\frac{\sinh(\frac{\pi k}{2})}  {2\cosh(\frac{\pi k}{2})+1}
  \frac{\delta(k-k_0)  r(k)}{\lambda-M(k)},
\eea
we see that if $\lambda<-1/3$ or $\lambda>0$, $\lambda-M(k)$ can not
have a zero to cancel the zero from $r(k)$ at  $k=k_0$ (recall that
$M(k)\in[-1/3,0)$ and $r(k_0)=0$). 
Therefore the integrations give zero and
$B_e=B_o=0$ and so ${\cal C}_e={\cal C}_o=0$.
Furthermore, $\delta(k-k_0)r(k)/(\lambda-M(k))$ will
be zero also. This means that $h(k)$ in \eref{v2_4} is zero.

Therefore in order to get nonzero $h(k)$ when $Det \ne 0$ we must 
require that $\lambda\in[-1/3,0)$ so that 
$\lambda-M(k)$ can cancel the zero coming from $r(k)$. In other words,
we find a {\bf continuous spectrum
$\lambda\in [-1/3,0)$}. Now we construct the eigenvectors for given 
$\lambda$. First we must 
choose the parameters $k_0$ and $r(k)$ such that 
\beq
\lambda=M(k_0) = -\frac{1}{1+2\cosh \frac{\pi k_0}{2}}
\label{k0_def} 
\eeq 
and $r(k)/(\lambda-M(k))$ is finite at $k=k_0$ 
(the $\lambda=-1/3$ case is a little more complex and we will discuss it
later). 
Knowing $k_0$ we can expand
\bea
\lambda-M(k)& = & M(k_0)-M(k)=-\frac{dM}{dk}|_{k_0} (k-k_0)-\frac{1}{2}
  \frac{d^2M}{dk^2}|_{k_0}(k-k_0)^2+....   \\
  & = & -\frac{\pi\sinh\frac{\pi k_0}{2} }{(1+2\cosh \frac{\pi k_0}{2})^2}
  (k-k_0)-\frac{1}{2} \frac{\pi^2+\frac{\pi^2}{2}\cosh \frac{\pi k_0}{2}
  -\pi^2 \sinh^2\frac{\pi k_0}{2} }{(1+2\cosh \frac{\pi k_0}{2})^3}
  (k-k_0)^2+....  \nonumber
\eea
For $k_0\neq 0$, $\frac{dM}{dk}|_{k_0} \neq 0$ so $r(k)$ can be chosen
as $D\cdot (k-k_0)$ where $D$ will be an overall normalization
constant and can be set to any value; we shall take $D=1$. 
Substituting into \eref{B_e_1} and
\eref{B_o_1}, we have
\bea
\label{B_e_2} B_e & = & -\frac{(\cosh(\frac{\pi k_0}{2})-1)
   (2\cosh(\frac{\pi k_0}{2})+1)}{\pi k_0 \sinh\frac{\pi k_0}{2}}, \\
\label{B_o_2} B_o & = & - \sqrt{3} \frac{(2\cosh(\frac{\pi
k_0}{2})+1)}{\pi k_0}. 
\eea 
Putting these results back into \eref{v2_4} we can get
\bean
\int_{-\infty}^{+\infty} dk h(k)\ket{k} & = & \int_{-\infty}^{+\infty} dk
\frac{4(\lambda-1)}{3\beta (1-\lambda)-2b} 
  \frac{\ket{k}\bra{k}v_e\rangle }{{\cal N}(k) (\lambda-M(k))}{\cal C}_e \\
 & &   + \int_{-\infty}^{+\infty}
dk\frac{4}{3\beta}\frac{\ket{k}\bra{k}v_o\rangle }{{\cal
N}(k)(\lambda-M(k))} {\cal C}_o 
    +\int_{-\infty}^{+\infty} dk \frac{\delta(k-k_0)
r(k)}{\lambda-M(k)}\ket{k}\\ 
& = &\frac{4(\lambda-1)}{3\beta (1-\lambda)-2b} {\cal C}_e \frac{1}{
  \lambda-M} \ket{v_e} + \frac{4}{3\beta} {\cal C}_o \frac{1}{
  \lambda-M} \ket{v_o}-\frac{1}{\frac{dM}{dk}|_{k_0}} \ket{k_0}.
\eean

We summarize the results as follows. For
every $\lambda\in[-1/3,0)$ we have two eigenvectors $v(k_0),v(-k_0)$
corresponding to the eigenvalue $\lambda = M(k_0) = -\frac{1}{1+2\cosh
\frac{\pi k_0}{2}}$:
\be
\label{eigen_v}
v(k_0)=\left[ \begin{array}{c} \frac{2\sqrt{2b}}{3\beta
(1-\lambda)-2b} {\cal C}_e(k_0) 
\\
\frac{4(\lambda-1)}{3\beta (1-\lambda)-2b} {\cal C}_e(k_0) \frac{1}{
  \lambda-M} \ket{v_e} + \frac{4}{3\beta} {\cal C}_o(k_0) \frac{1}{
  \lambda-M} \ket{v_o}-\frac{1}{\frac{dM}{dk}|_{k_0}} \ket{k_0}
\end{array}  \right].
\ee
As mentioned in the introduction,
there is a subtlety when $b > b_0 := 8 \ln 2$, here the forms of
\eref{eigen_v} become modified at one single point. 
From this expression of the eigenvectors, we see that
the eigenvector of $M'$ can be seen as a deformation of that 
of $M$ at $\ket{k_0}$ by a proper linear combination of $\ket{v_e}$ 
and $\ket{v_o}$. This is a special property for the continuous
spectrum.
As we will see, for the discrete spectrum, they are just the linear
combinations of $\ket{v_e}$ and $\ket{v_o}$ without involving $\ket{k_0}$.

Notice that since for every $\lambda$ we have doubly degenerate
eigenvectors 
$v(k_0),v(-k_0)$, we can  use the relations
\bean
{\cal C}_e(k_0)={\cal C}_e(-k_0),~~~~~~~~{\cal C}_o(k_0)=-{\cal C}_o(-k_0),
~~~~~~~\frac{dM}{dk}|_{k_0}=-\frac{dM}{dk}|_{-k_0}
\eean
to construct a {\bf twist even} eigenstate
\be
\label{v_even}
v_{+}=\frac{1}{2}(v(k_0)+v(-k_0))=\left[ \begin{array}{c}
\frac{2\sqrt{2b}}{3\beta (1-\lambda)-2b} {\cal C}_e(k_0) 
\\
\frac{4(\lambda-1)}{3\beta (1-\lambda)-2b} {\cal C}_e(k_0) \frac{1}{
  \lambda-M} \ket{v_e} -\frac{1}{2\frac{dM}{dk}|_{k_0}}
(\ket{k_0}-\ket{-k_0})
\end{array}  \right]
\ee
as well as a {\bf twist odd} eigenstate
\be
\label{v_odd}
v_{-}=\frac{1}{2}(v(k_0)-v(-k_0))=\left[ \begin{array}{c} 0 \\
\frac{4}{3\beta} {\cal C}_o(k_0) \frac{1}{
  \lambda-M} \ket{v_o}-\frac{1}{2\frac{dM}{dk}|_{k_0}} (\ket{k_0}+\ket{-k_0})
\end{array}  \right].
\ee
We remind the reader that $k_0$ is defined in \eref{k0_def}. 
Also ${\cal C}_e,{\cal C}_o$
can be found from \eref{Ce_1}, \eref{Co_1} 
and \eref{B_e_2}, \eref{B_o_2}. Finally
$\frac{dM}{dk}=\frac{\pi\sinh\frac{\pi k_0}{2} }{(1+2\cosh \frac{\pi
k_0}{2})^2}$. 
\section{The Determinant: the Functions $A_{ee}$ and $A_{oo}$}
We have seen from the setup that to completely determine the
eigenvectors and eigenvalue of $M'$ we must understand
the behavior of the determinant $Det$ in the linear system
\eref{v2_8}. It is therefore crucial to first understand
the behaviour of $A_{ee}$ and $A_{oo}$ as functions of $\lambda$.
We will give the analytic forms of these functions, analyze their
singularities and find the critical $\lambda$'s which make $Det$
zero.
\subsection{The Function $A_{ee}$}
By summing all the residues in the upper-half plane, one can
analytically evaluate the integral $A_{ee}(\lambda)$, which we recall
from \eref{ab} as
\[
A_{ee}(\lambda) = 
\int_{- \infty}^{\infty} {dt \over t} {\sinh(t/2)^2 \tanh(t/2) \over 
\left( 1 + 2 \cosh(t) \right) \left(1 + \lambda + 2 \lambda 
\cosh(t) \right)}.
\]
The result is
\bea
 A_{ee}(\lambda) &=& 
 \frac{-1}{4 (\lambda-1)} \left\{ \bit{-0.1}
9 (\lambda-1) \ln3 + 2 ( \gamma + 
3 \gamma \lambda + 8 \ln2) \right. \nonumber \\ 
&+& \left. (1 + 3 \lambda) \left( \bit{-0.1}
\psi[-g(\lambda)] + 
\psi[\alpha(\lambda) + g(\lambda)] \right) \right\}, \nonumber
\\
\eea
where $\gamma$ is Euler's constant, and $\psi(z)$ is the 
{\em digamma function}
$\psi(z) = {d \over dz} \ln \Gamma(z)$. Furthermore,
\be
\label{gash_it}
g(\lambda) := {i \over 2 \pi} \Arcsech\left(-{2 \lambda \over 1 +
\lambda} \right) \qquad {\rm and} \quad
\alpha(\lambda) := \left\{
\begin{array}{cc}
1, & \lambda \notin (-\frac13,0)\\
0, & \lambda \in (-\frac13,0) \; .
\end{array}
\right.
\ee
For reference, we plot $A_{ee}$ in Figure \ref{aee}. 
\EPSFIGURE[h]{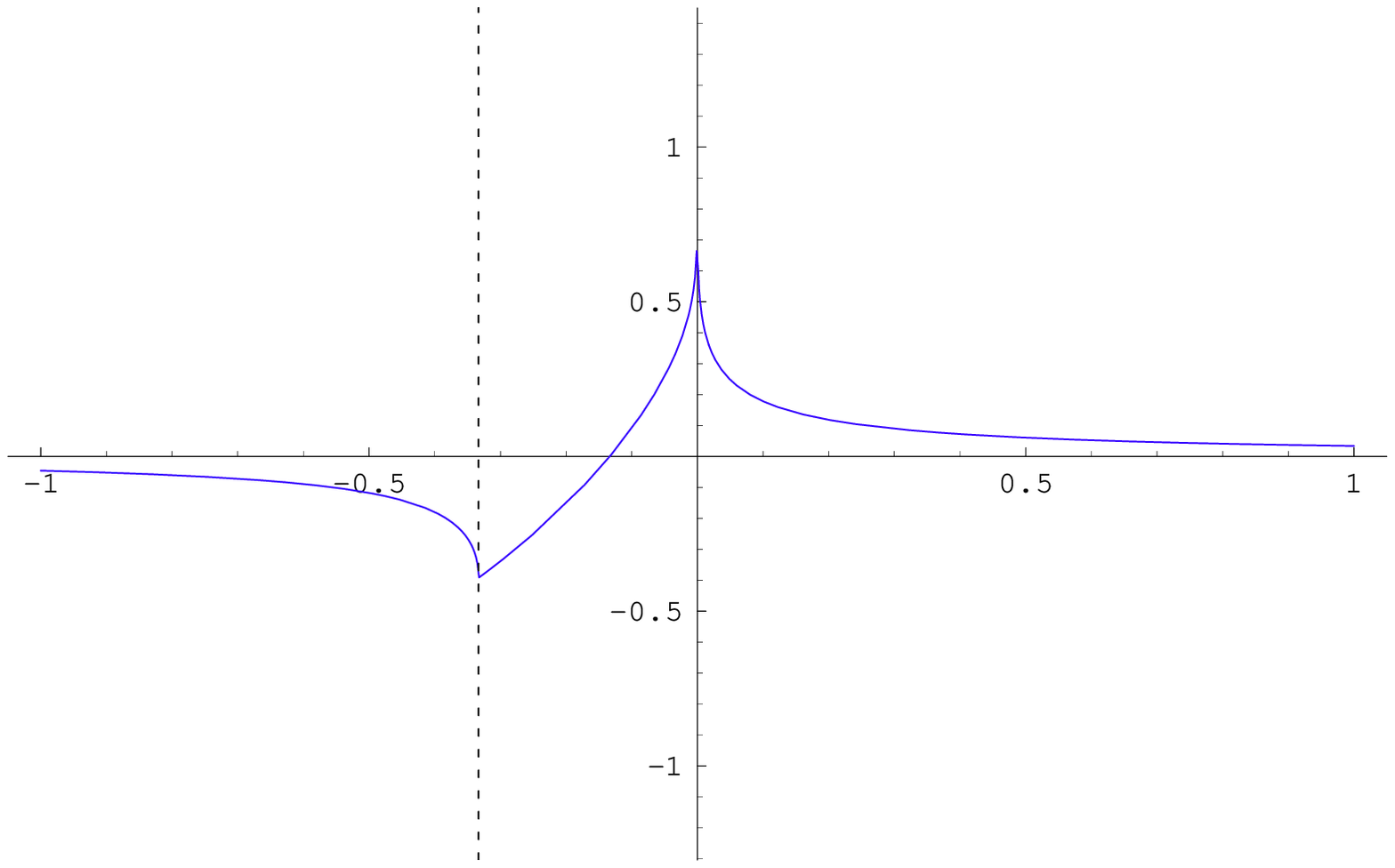,width=15cm}
{$A_{ee}$ as a function of $\lambda$. The dashed line is at $\lambda =
-1/3$.
\label{aee}
}
Let us note a few key features. 
It seems that when $\lambda=1$, $A_{ee}$ is not
well defined. However, 
careful analysis will show that in fact $A_{ee}$ is continuous there
and
$$
A_{ee}(\lambda=1)=-\frac{3}{4} V_{00}^{rr} + \frac{7\zeta(3)}{2\pi^2}
$$
where $\zeta(z)$ is the celebrated {\em Riemann $\zeta$-function}.

Also despite the discontinuity of $\alpha(\lambda)$,
$A_{ee}$ is well-defined at $\lambda=-1/3=M(k=0)$. We can compute both
limits from the left and the right to obtain
\be
\label{aee-1/3}
A_{ee}(-\frac13)=-\frac{3}{4}V_{00}^{rr} = -\frac34\ln\frac{27}{16}. 
\ee
This incidentally proves the identity \eref{Spec_Okuyama_2_3.13},
which has so far escaped the literature\footnote{This is due to the
	fact that from \eref{Spec_Okuyama_2_3.13}, we have the
expression for 
	$A_{ee}$ at $\lambda=-1/3$ as
\[
A_{ee}(\lambda = -\frac13)
= \int_{-\infty}^{+\infty} dk
  \frac{\bra{k}v_e\rangle \langle v_e\ket{k}}{{\cal N}(k) (-1/3-M(k))}
= -3 \int_{-\infty}^{+\infty} dk
  \frac{\bra{k}v_e\rangle \langle v_e\ket{k}}{{\cal N}(k) (1+3M(k))}
= -3 \bra{v_e}\frac{1}{1+3M}\ket{v_e}.
\]
	}.
The reason for this good behaviour is that 
near $k=0$, ${\cal N}(k)\sim 1$, $\bra{k}v_e\rangle \sim k$
and $\lambda-M(k)=-1/3-M(k) \sim k^2$, so the integrand is 
well defined. This is not true for $\lambda=0$ where $A_{ee}$ diverges
as $\ln\ln(\lambda)$. In fact
\[
A_{ee}(\lambda \rightarrow 0) \sim 
\frac{2\,\gamma + 16\,\ln (2) - 9\,\ln (3) - 2\,\ln (2\,\pi ) +
2\,\ln (-\ln \abs{\lambda})}{4}
\]

One root of $Det$ can be found by solving
\be
\label{implicit}
A_{ee}=-\frac{3}{4}V_{00}^{rr}+b\left(
\frac{1}{2(1-\lambda)}-\frac{3}{8} \right)\equiv 
I_b(\lambda)
\ee
By studying the intersection of $I_b(\lambda)$ with $A_{ee}(\lambda)$
we see that there are two kinds of roots (cf. Figure \ref{aeeIb}). 
We note that $I_b(\lambda)$
is a hyperbola with asymptote at $\lambda=1$ so from $-\infty$ to $1$
it is an increasing function from $-\frac34 \beta$  to
$\infty$.  Therefore
the first kind of root exists no matter what $b$ is (we recall that
$b > 0$), namely 
they are $\lambda=-1/3$ (because $I_b$ always passes through the point
$(-1/3, -3/4 \ln(27/16)) \sim (-1/3, -0.392436)$, 
the left cusp point of $A_{ee}$; we will show this below) and 
some $\lambda_1 \in (0,1)$. However when $I_b$ increases fast enough,
it could intersect $A_{ee}$ one more time in the region $[-1/3,0)$;
this is when
$\frac{dI_b}{d\lambda}|_{-1/3} \ge \frac{dA_{ee}}{d\lambda}|_{-1/3}$. So
the critical point occurs at $\frac{dI_b}{d\lambda}|_{-1/3} =
\frac{dA_{ee}}{d\lambda}|_{-1/3} \Rightarrow b = 8\ln 2$. Therefore
a second kind of root exists in addition to the first only when $b\geq
8\ln 2$ and is located in the region $[-1/3,0)$.

As promised, we will now show 
that indeed $\lambda=-1/3=M(k=0)$ gives
$1-\frac{4(\lambda-1)}{3\beta (1-\lambda)-2b}A_{ee}=0$. To see this,
we recall from \eref{aee-1/3} that
$A_{ee}(\lambda = -\frac13) =  -\frac{3}{4} V_{00}^{rr}$.
Using this we can calculate 
$$
1-\frac{4(\lambda-1)}{3\beta (1-\lambda)-2b}A_{ee} = 
1-\frac{4(-1/3-1)}{3(V_{00}^{rr}+b/2)(1+1/3)-2b} (-\frac{3}{4}) V_{00}^{rr} 
 =  0.
$$

We see therefore that $I_b(\lambda)$ passes through the left cusp of
$A_{ee}(\lambda)$ and $\lambda = -\frac13$ indeed is a root of $Det$.
\EPSFIGURE[ht]{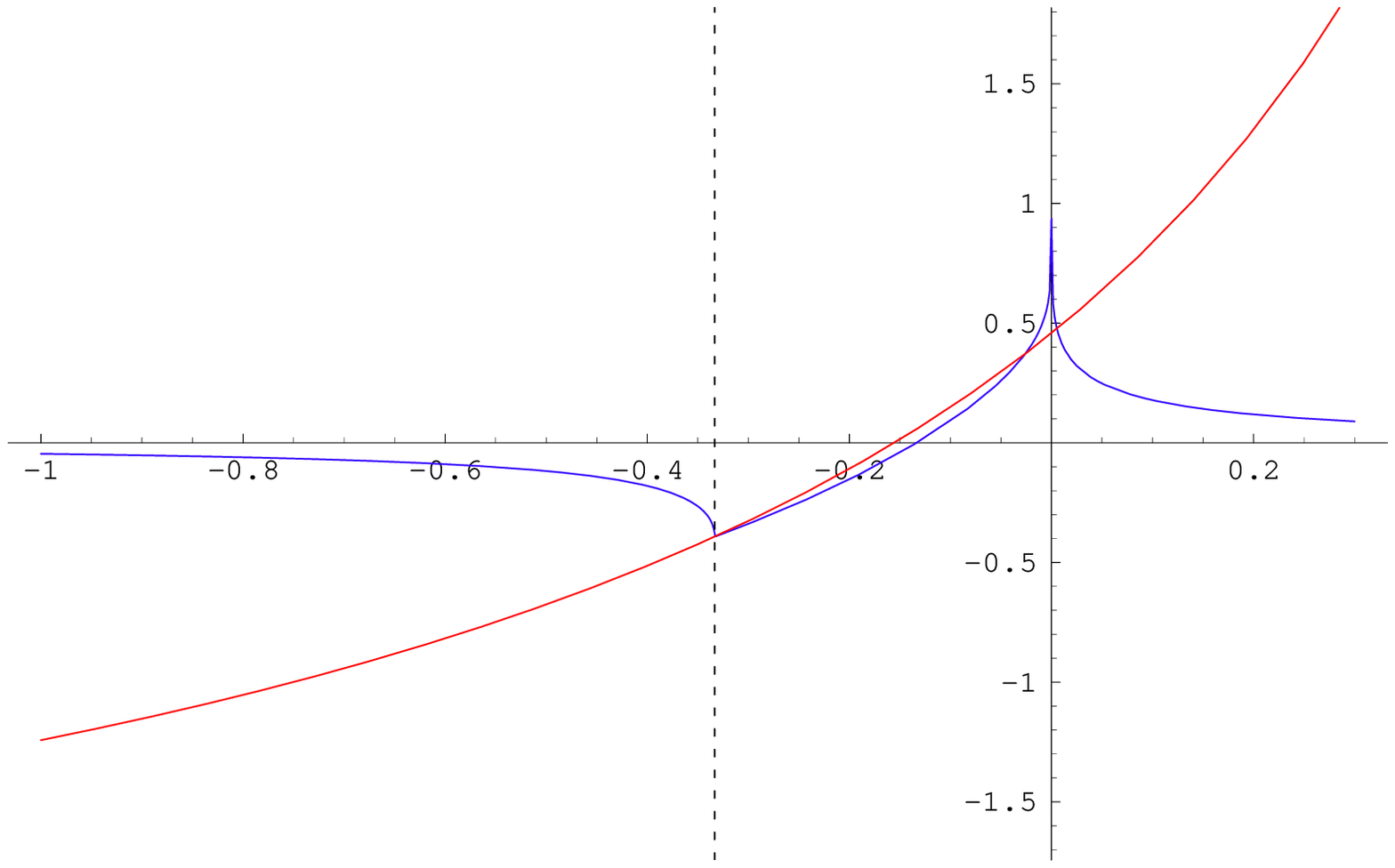,width=15cm}
{The intersection of $A_{ee}$ with $I_b$ as functions of $\lambda$. We
have here chosen a $b$ above the critical value $b_0$ so that we can
explicitly see 3 points of intersection. Note that both $A_{ee}$ and
$I_b$ always meet at least at the dashed line at $\lambda = -1/3$.
\label{aeeIb}
}
\subsection{The Function $A_{oo}$}
By the same method we can evaluate 
\[
A_{oo}(\lambda) := \int_{-\infty}^{\infty}{dt \over t}{3 \sinh(t)
\over 2 \left(1+2 \cosh(t)
\right) \left(1 + \lambda + 2 \lambda \cosh(t) \right)}.
\]
Now we obtain
\[
A_{oo}(\lambda) = 
{3 \over 4} \left( \bit{-0.1} 2 \gamma + 3 \ln 3 + 
\psi\left[ 1+ g(\lambda)\right] + 
\psi\left[1 - \alpha(\lambda)-g(\lambda)\right]\right),
\]
where $g(\lambda)$ and $\alpha(\lambda)$ were defined in \eref{gash_it}.
\EPSFIGURE[h]{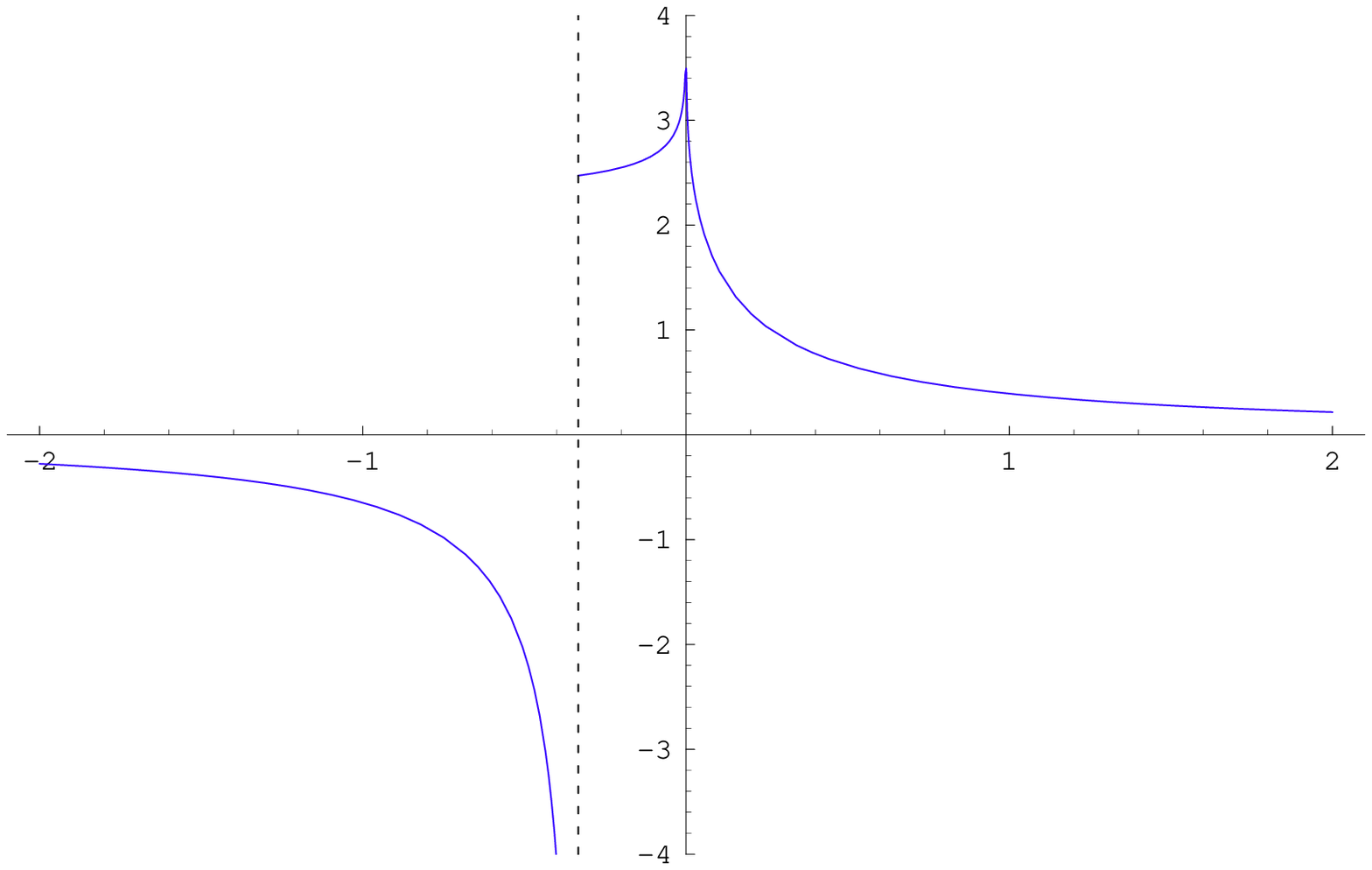,width=15cm}
{$A_{oo}$ as a function of $\lambda$. The dashed line is at $\lambda = -1/3$.
\label{aoo}
}
We plot $A_{oo}$ in Figure \ref{aoo}. There are several 
important points here as well. Firstly putting $\lambda=1$ we get
$\langle v_o|\frac{1}{1-M}\ket{v_o}=\frac{3}{4} V_{oo}^{rr}$, giving
us the nice identity.
Secondly $A_{oo}$ diverges at $\lambda=0$ from both
sides. The divergence is again very slow, as $\ln\ln\lambda$:
\[
A_{oo}(\lambda \rightarrow 0) \sim \frac{3\,\left(
  2\,\gamma + 3\,\ln (3) - 2\,\ln (2\,\pi ) \right)
  }{4} +  
  \frac{3\,\ln (-\ln (\abs{\lambda}))}{2}
\] 
More important is the behavior
near $\lambda=-1/3$. If we approach from the left
we find $A_{oo}|_{(-\frac{1}{3})^-}=-\infty$. If we approach
from the right, we find $A_{oo}|_{(-\frac{1}{3})^+}=\frac{3}{4}\ln 27$ 
which is finite. This discontinuity may seem unnatural,
but we will see later that it is consistent with our analysis.

Now we can solve the other $\lambda$ which makes $Det$ zero. 
The equation is
\be \label{a_oo_det}
A_{oo}=\frac{3\beta}{4} = \frac34 \left(\ln \frac{27}{16} + \frac{b}{2}\right).
\ee
From this we find again that there are two kinds of solutions. The first
one does not depend on the value of $b$ and is located in the region
$(0,1)$ (since $b > 0$). For large enough $b$, of course, we obtain
a second type of zero in addition to the first, located in the region
$[-1/3,0)$. This occurs when the right hand side is higher than when
$A_{oo}$ takes its point of discontinuity at $\lambda = -1/3$; this is
when $b \ge 8\ln 2$.
Comparing with the critical value of $b$ found in the
$A_{ee}$ case, we find they are same. This is not an accident. 

In fact
we claim that the solutions $\lambda$ found in both cases, either from
$A_{ee}$ or from $A_{oo}$, whether
in the region $(0,1)$ or $[-1/3,0)$ are the same, i.e., the two roots
of $Det$ are {\bf degenerate}.
To show this, we use the
analytic form of  $A_{ee}$ and $A_{oo}$, giving the ratio
\be
\label{ratio_out}
\frac{1-\frac{4}{3\beta} A_{oo}}{1-\frac{4(\lambda-1)}{3\beta
(1-\lambda)-2b}A_{ee}}=\frac{ b + 3 b \lambda + 6 (\lambda-1)
\ln\left({27 \over 16}\right)}
{(1+3\lambda) \left(b+2\ln\left({27 \over 16}\right)
\right)}.
\ee
As analysed, the denominator gives one root of $Det$ and the
numerator, the other.
The idea is that if the roots are degenerate, they
will cancel each other so that this ratio is neither zero nor
infinite at the roots.
From
\eref{ratio_out} we see that the ratio is zero only when 
$\lambda=(6\ln(27/16)-b)/(3b+6 \ln(27/16))$; careful analysis reveals
that this zero is coming from the simple pole in the denominator at
$3\beta (1-\lambda)-2b=0$ and so is in fact not a zero of $Det$. On
the other hand, the only pole is at $\lambda = -1/3$.
We hence conclude that the two zeros of $Det$ are
degenerate\footnote{The degeneracy between the zeros of $A_{oo}$ and $A_{ee}$
	is broken in the limit $b=0$. In this case, $Det=0$ 
	for $A_{oo}$ has solution at $\lambda=1$ while there is no solution
	for $A_{ee}$ in the region $(0,1]$. However this case of $b=0$
	is not a physical choice.} except for $\lambda = -\frac13$
which is a zero of the denominator only (for all values of $b$).
\section{The Discrete Spectrum}
Having discussed the continuous spectrum, we now move on to the
discrete spectrum.  This comes from the zeros of
the determinant $Det$. The solutions have been discussed in section
5. In this section, we will construct the corresponding eigenvectors.
\subsection{The Case of $\lambda=-1/3$}
As we have shown, no matter what $b$ is, 
$M_{ee}= 1-\frac{4(\lambda-1)}{3\beta (1-\lambda)-2b}A_{ee}=0$
always has a solution $\lambda=-1/3$. We will denote the corresponding
eigenvector as $v_{+,-\frac{1}{3}}$. Furthermore, when $b\geq 8\ln2$,
both $M_{ee}$ and $M_{oo}=1-\frac{4}{3\beta} A_{oo}$ have another
solution $\lambda\in[-1/3,0)$. When $b= 8\ln 2$, the solution will be
$\lambda=-1/3$ again, which is also 
degenerate\footnote{Notice that the existence of zeros for $M_{oo}$ 
	at $b= 8\ln 2$ depends crucially on the existence of the limit of
	$A_{oo}$ when we reach $\lambda=-1/3$ from the right.}. 

Now we can start to construct the eigenvectors. Since $M_{ee}=0$,
for consistency of \eref{v2_8}, we need $B_e=0$. 
This can be achieved by choosing any
$k_0 \ne 0$ or $k_0=0$ such that $r(k=0)/(-\frac13-M(k=0))$ is not a pole. 

If we choose $k_0\neq 0$, we have $B_o=0$ and the solution is\footnote{
	Here in principle we can choose ${\cal C}_e$ to be any non-zero
	value. What we choose here is just a convenience to compare the
	result with \cite{Spec_Oku_2}.}
\be
{\cal C}_e=\frac{2V_{oo}^{rr}}{\sqrt{2b}},~~~~~~~{\cal C}_o=0
\ee
and the eigenvector becomes
\bean
g & = & 1, \\
\int_{-\infty}^{+\infty} dk h(k) \ket{k} & = &\int_{-\infty}^{+\infty} dk
  \left(-\frac{8}{3 \sqrt{2b}} \frac{\ket{k}\bra{k}v_e \rangle}{{\cal N}(k)
	(-1/3-M(k))} \right)\\ 
& = & -\frac{8}{3 \sqrt{2b}}\frac{1}{ (-1/3-M)}\ket{v_e}  \\
& = & \frac{8}{ \sqrt{2b}}\frac{1}{ 1+3M}\ket{v_e}.
\eean
In summary then,
\be
\label{1/3-}
v_{+,-\frac{1}{3}}=\left[ \begin{array}{c} 1 \\
\frac{8}{ \sqrt{2b}}\frac{1}{ 1+3M}\ket{v_e}  \end{array} \right]
\ee
which is the solution given in \cite{Spec_Oku_2} (equation (4.5)). Notice 
that this state is {\em twist-even}. This solution
has been found by several groups already \cite{MT,Spec_Oku_2,Barton}.

If instead of choosing $k_0\neq 0$, we choose 
$k_0=0$ such that $r(k)/(-1/3-M(k))$ does not have a pole at $k=0$, then
there are 
two cases. The first one is that $r(k)/(-1/3-M(k))$ has a zero at $k=0$,
so we have $B_o=0$ and the solution is the same as above. The second one is
that $r(k)/(-1/3-M(k))\sim 1$  at $k=0$, then we will have a non-zero
$B_o$. We point out that this is when $b \ne b_0 := 8 \ln 2$. Indeed
if $b =  b_0$, consistency of \eref{v2_8} requires $Det$ and hence
$B_o$ to be zero.
This non-zero
$B_o$ opens the possibility for another eigenvector. If we choose the branch
$A_{oo}|_{(-\frac{1}{3})^{-}}=-\infty$, we will have ${\cal C}_o=0$ although
$B_o\neq 0$. However, if we choose the branch 
$A_{oo}|_{(-\frac{1}{3})^{+}}=\frac{3}{4}\ln 27$, we get a nonzero
${\cal C}_o$.
In this case we can construct two eigenvectors: one is twist-even and
one is twist odd. 

Let us work out the details. Setting $k_0=0$ and expanding around
$k=0$ we obtain $(-1/3-M(k)) \sim k^2 + O(k^3)$ (the first order is
zero). Therefore we can choose the parametre $r(k)=k^2$ and
get  $B_o=-6\sqrt{3}/\pi$. Then ${\cal C}_o=-\frac{6\sqrt{3}
}{\pi(1- \frac{\ln 27}{\beta})}$. 
If we set ${\cal C}_e=0$, we get the eigenvector as
\be
\label{1/3+-}
v_{-,-\frac{1}{3}}=\left[ \begin{array}{c} 0 \\
\frac{4{\cal C}_o}{3\beta} \frac{1}{-1/3-M}\ket{v_o}-\frac{36}{\pi^2}
\ket{k=0}\end{array} \right].
\ee
We can check this directly by acting $M'$ on the left. Using
$$
\bra{v_e}k=0\rangle=0, \quad \bra{v_e} \frac{1}{-1/3-M} \ket{v_o}=0,
\quad \bra{v_o}k=0 \rangle=\frac{\sqrt{3} \pi}{6}, \quad
A_{oo}|_{(-\frac{1}{3})^{+}}=\frac{3}{4}\ln 27.
$$
If we choose ${\cal C}_e=\frac{2V_{oo}^{rr}}{\sqrt{2b}}$, we will
get 
$$
v'=\left[ \begin{array}{c} 1 \\\frac{8}{ \sqrt{2b}}\frac{1}{
1+3M}\ket{v_e} +
\frac{4{\cal C}_o}{3\beta} \frac{1}{-1/3-M}\ket{v_o}-\frac{36}{\pi^2}
\ket{k=0} \end{array} \right].
$$
From these two solutions we can construct the twist-odd solution 
$v_{-,-\frac{1}{3}}$ and the twist-even solution 
$v_{+,-\frac{1}{3}}=v'-v_{-,-\frac{1}{3}}$, which is equal to \eref{1/3+-}.
In fact, comparing with the results \eref{v_even} and \eref{v_odd} 
from the last section, we find that
these two solution $v_{\pm,-\frac{1}{3}}$ are nothing new, but a part
of the continuous spectrum we presented before.  

It is a little strange that we get twist-even and twist-odd states
for $M'$ at $\lambda=-1/3$ at the same time while for $M$, we have
only a 
twist-odd state. To see that it is true, let us take $b\rightarrow
+\infty$.  In this limit we have from \eref{Spec_Okuyama_2_3.1}
$$
M'=\left[ \begin{array}{cc} -\frac{1}{3} & 0 \\ 0 & M \end{array}
\right]. 
$$
From this limit, we see immediately that $M'$ has two eigenvectors for
the eigenvalue $-\frac{1}{3}$:
$$
v_{+}=\left[\begin{array}{c} 1  \\ 0  \end{array} \right],~~~~~~
v_{-}=\left[\begin{array}{c} 0  \\ \ket{k=0}  \end{array} \right];
$$
these are of course nothing other than the limit of
$v_{\pm,-\frac{1}{3}}$ when $b\rightarrow +\infty$. We consider this
as a strong evidence supporting the double degeneracy at $\lambda =
-1/3$. In the conclusions section, we will give some numerical
evidence and further discussion about this point.

We have discussed the case of $b \ne b_0$ in the above and found that
the discrete spectrum at $\lambda = -1/3$ is the same as the
continuous at this point.
Now we discuss the special case when $b = b_0=8\ln2$. 
Recall from Subsection 5.2, we must choose the branch of  
$A_{oo}|_{(-\frac{1}{3})^{+}}=\frac{3}{4}\ln 27$ in order to get a
zero for the determinant. Consistency of \eref{v2_8} requires
$B_e=B_o=0$. This can be achieved by setting $k_0\neq 0$ or by setting
$k_0=0$ but with $r(k)/(-1/3-M(k))$ having a zero at $k=0$.
In either choice we will get two eigenvectors by letting 
 ${\cal C}_e\neq 0,~~{\cal C}_o=0$
or ${\cal C}_e= 0,~~{\cal C}_o\neq 0$. The results are
\be
\label{v_even_b0}
v^{b_0}_{+,-\frac{1}{3}}=\left[ \begin{array}{c} 1 \\
\frac{8}{ \sqrt{2b}}\frac{1}{ 1+3M}\ket{v_e}  \end{array} \right]
\ee 
and
\be
\label{v_odd_b0}
v^{b_0}_{-,-\frac{1}{3}}=\left[ \begin{array}{c} 0 \\
\frac{4{\cal C}_o}{3\beta} \frac{1}{-1/3-M} \ket{v_o} \end{array}
\right].
\ee
Notice that although $v^{b_0}_{+,-\frac{1}{3}}$ is the same as
\eref{1/3-}, $v^{b_0}_{-,-\frac{1}{3}}$ is different from \eref{1/3+-}
by missing  the 
$\ket{k=0}$ term. This is a very important point. It in fact
distinguishes the 
continuous and the discrete spectra. This means that 
when $b_0 \ne 8\ln2$, the continuous spectrum at $\lambda=-1/3$ is 
simply the discrete spectrum at this point. However when $b_0
= 8\ln2$, the expressions \eref{v_even} and \eref{v_odd} for the
continuous spectrum at $\lambda=-1/3$  no longer apply but should be
replaced by \eref{v_even_b0} and \eref{v_odd_b0}.
\subsection{Other Solutions at $\lambda\neq -1/3$}
For other $\lambda\neq -1/3$ which make $Det$ zero no matter which
region they are, the eigenvectors can be found similarly. First we 
choose ${\cal C}_e\neq 0,~~{\cal C}_o=0$ and the eigenvector
is twist even  
\be \label{iso_even}
v_{+}=\frac{1}{2}(v(k_0)+v(-k_0))=\left[ \begin{array}{c}
\frac{2\sqrt{2b}}{3\beta (1-\lambda)-2b} {\cal C}_e(k_0) 
\\
\frac{4(\lambda-1)}{3\beta (1-\lambda)-2b} {\cal C}_e(k_0) \frac{1}{
  \lambda-M} \ket{v_e} 
\end{array}  \right].
\ee
Next we choose ${\cal C}_e= 0,~~{\cal C}_o\neq 0$ and the eigenvector
is twist odd
\be \label{iso_odd}
v_{-}=\frac{1}{2}(v(k_0)-v(-k_0))=\left[ \begin{array}{c} 0 \\
\frac{4}{3\beta} {\cal C}_o(k_0) \frac{1}{
  \lambda-M} \ket{v_o}
\end{array}  \right].
\ee
Again, when $\lambda\in [-1/3,0)$ the expressions \eref{v_even} and
\eref{v_odd} for the continuous spectrum will
be replaced by these above expressions for the discrete spectrum.
\section{The Generating Function}
In the above sections, we have given the eigenvectors of $M'$ for the
various ranges of $\lambda$. They are of the form of
$\ket{v_e}$ and $\ket{v_o}$ acted on by ${1 \over \lambda-M}$. It
would be very nice if we could explicitly determine these components.
The present section solves this problem.

In order to find components, we need to find the generating function.
The idea is that, recalling $f_k$ in \eref{f_gen} we can define
generating functions $G_e(z)$ and $G_o(z)$ as follows:
\bea
G_e(z)\equiv\bra{z} E^{-1} \frac{1}{\lambda-M} \ket{v_e} & = &
\int_{-\infty}^{+\infty} 
  dk { f_k(z) \bra{k} v_e \rangle \over {\cal N}(k)(\lambda-M(k)) },
\label{gen_ve}\\
G_o(z)\equiv\bra{z} E^{-1} \frac{1}{\lambda-M} \ket{v_o} & = &
  \int_{-\infty}^{+\infty} 
  dk { f_k(z) \bra{k} v_o \rangle \over {\cal N}(k)(\lambda-M(k)) }.
\label{gen_vo}
\eea
The series expansion coefficients in $z$ of $G_e(z)$ (respectively
$G_o(z)$) will give the components of ${1 \over \lambda-M}\ket{v_e}$
(respectively  ${1 \over \lambda-M}\ket{v_o}$).

Recalling the definition of $f_k$ in \eref{f_gen}, as well as
\eref{f_gen} and \eref{f_gen_1}, in addition to \eref{M_eig} and
\eref{inner}, the integrals have the explicit forms
\bea
G_e(z) &=& \int_{-\infty}^{+\infty} dk
	\frac{\left( 1 - e^{- k \arctan z} \right) \,
    \left( -1 + \cosh (\frac{k\,\pi }{2}) \right) \,
    }{2\,k\,
    \left( 1 + 2\,\cosh (\frac{k\,\pi }{2}) \right) \,
    \left( \lambda + 
      \frac{1}{1 + 2\,\cosh (\frac{k\,\pi }{2})} \right)
	\sinh(\frac{k\,\pi }{2})}\\ 
G_o(z) &=& \int_{-\infty}^{+\infty} dk
\frac{{\sqrt{3}}\,\left( 1 - e^{-k \arctan z }
      \right) }{2\,k\,\left( 1 + 
      2\,\cosh (\frac{k\,\pi }{2}) \right) \,
    \left( \lambda + 
      \frac{1}{1 + 2\,\cosh (\frac{k\,\pi }{2})} \right) }
\eea
Our task is therefore to evaluate the above two integrals. Again
summing up the residues on the upper half plane we obtain the
following. 
\subsection{The Twist-even States}
For the generating function $G_e(z)$, when $\lambda \in[-1/3,0)$,
setting $\lambda=-(2\cosh(\frac{\pi k_0}{2})+1)^{-1}$ we have\footnote{All
	ensuing results will be correct only for $|z| < \frac{\pi}{4}$
	because of a choice of branch cut; this is no hindrance
	because $z$ is merely an expansion parametre.}
\bea
\hspace{-1cm}
G_e(z) & = & {1+2\cosh( { \pi k_0 \over 2}) \over 4 k_0 (1+\cosh( {
\pi k_0 \over 2}))} \left(
 k_0 B[e^{-4 i \arctan z}; 1-{ i
k_0 \over 4},0]+ k_0 B[e^{-4 i \arctan z}; 1+{ ik_0 \over 4},0] \right.
\nonumber \\
& & \left. + k_0 (2 \gamma -4 \Arctanh(e^{- 2 i \arctan z}) + 
\ln(16)+\psi(-{ ik_0 \over 4})+ \psi({ ik_0 \over 4}) )
-4 i \sinh(k_0 \arctan z) \right)
\eea
where
\[
B[z;a,b]\equiv B_z[a,b]=\int\limits_{0}^z t^{a-1} (1-t)^{b-1}
dt
\] 
for $(\Re(a)>0)$ is the {\em incomplete beta function}.

For $\lambda_1 \in (0,1)$ we set 
\beq
\lambda_1=(2\cosh(\frac{\pi k_0}2)-1)^{-1}
\eeq
and have
\bea
G_e(z) & = & { i (-1+2\cosh( { \pi k_0 \over 2})) \csch^2( { \pi k_0
\over 4}) \over 2}\left( i \Arctanh(e^{-2i \arctan z})
\begin{array}{c} \\ \\ \end{array}
\right.
\nonumber \\
&&
-{i \over 4} \left( 2 \gamma + \ln(16) +\psi({1\over2}-{ ik_0 \over 4})+
\psi({1\over2}+{ik_0 \over 4})\right)
\nonumber \\
&&
+{e^{-(k_0+2 i) \arctan z}\over 2 i+k_0}
 ~_2F_1[{1\over2}-{ ik_0 \over 4}, 1, {3\over2}-
{ ik_0 \over 4}, e^{-4i \arctan z}]
\nonumber \\
&&
\left.
+{e^{(k_0-2 i) \arctan z}\over 2 i-k_0}
 ~_2F_1[{1\over2}+{ ik_0 \over 4}, 1, {3\over2}+
{ ik_0 \over 4}, e^{-4i \arctan z}]
\right)
\eea
where 
\[
_2F_1[a,b,c,z]={ \Gamma(c) \over
\Gamma(b) \Gamma(c-b)} \int\limits_0^1 t^{b-1} (1-t)^{c-b-1}
(1-tz)^{-a} dt
\]
(for $\Re(c)>\Re(b)>0;|{\rm Arg}(1-z)|\leq \pi$) is the {\em
hypergeometric function} of the first kind.

As an application of the above generating function, we derive the
components of the state $v_{+,-{1 \over 3}}$ in \eref{1/3+-}. Taking the
limit $\lambda\rightarrow -1/3$ (or equivalently $k_0\rightarrow 0$) 
we can simplify the generating function $G_e(z)$ as
\bea
G_e(z)|_{k_0=0} & = & {3 \over 4} \ln(1 + z^2) \\
& = & -{ 3 \over 4}\sum_{n\geq 1} \frac{(-)^{n}}{n} z^{2n} = -{3\over
2} \sum_{k=even} \frac{1}{\sqrt{k}} \frac{(-)^{k/2}}{\sqrt{k}} z^k.
\eea
We conclude therefore that (up to an overall factor) the twist-even
eigenvector at $\lambda = -1/3$ (from \eref{1/3-}) has components
\[
v_k = \frac{4}{\sqrt{2b}}\frac{(-1)^{k/2}}{\sqrt{k}} \quad \mbox{$k$
even and } > 0,
\]
$v_0 = 1$ and $v_k = 0$ for $k$ odd.
This reproduces the result given in equation (4.15) of
\cite{MT} for\footnote{We think that Equation (4.2) (and therefore
(4.15)) of
\cite{MT} is compatible with $b=4$, as can be checked by solving these
equations for $U'$. However their equations (2.3) and (2.9) seem to be
compatible with $b=2$. We think this might be an inconsistency between
equations (2.3, 2.9) and (4.2) of \cite{MT}.} $b=4$.
\subsection{The Twist-odd States}
Now we discuss the generating function $G_o(z)$. Again, 
when $\lambda \in[-1/3,0)$
we can set $\lambda=-1/(2\cosh(\pi k_0/2)+1)$ and obtain
\bea
G_o(z) & = & {\sqrt{3}(2 \coth({ \pi k_0 \over 2})+ \csch({ \pi k_0
\over 2})) \over 4} \left(B[e^{-4 i \arctan z}; 1-{ ik_0 \over 4},0]
- B[e^{-4 i \arctan z}; 1+{ ik_0 \over 4},0] \right. \nonumber \\
& & \left.  \qquad \qquad
+{ 4i \cosh(k_0 \arctan z) \over k_0}- i\pi \coth({ \pi
k_0 \over 4}) \right).
\eea
On the other hand, when $\lambda_1 \in (0,1)$ we can set
$\lambda_1=1/(2\cosh(\pi k_0/2)-1)$ and obtain
\bea
\hspace{-1cm}
G_o(z) &=& \frac{i \sqrt{3} (-1+2\cosh( { \pi k_0 \over 2}))
\csch({k_0 \pi \over 2})}
{8 k_0} \left( -i (k_0 - 2 i) B[e^{-4 i a},
{1\over2}-{i k_0 \over 4}, 0] \right.
\nonumber \\
&& + e^{-2 (4 i+k_0) a} \left( {4 e^{(6 i + k_0) a} (-2i+k_0+
2e^{2k_0 a}k_0) \over k_0-2i} ~_2F_1[1, {1\over2}+{ik_0 \over 4},
{3\over2}+{ik_0 \over 4}, e^{-4 i a}] \right.
\nonumber \\
&& -2 e^{(k_0+2i) a} \left( {8 i \over k_0 + 6 i}
~_2F_1[2, {3\over2}-{i k_0 \over 4}, {5\over2}-{i k_0 \over 4}, e^{-4 i
a}]
\right. \nonumber \\
&& \left. \left. \left.
+{8 i \over k_0 - 6 i} ~_2F_1[2, {3\over2}+{i k_0 \over 4},
{5\over2}+{i k_0 \over 4}, e^{-4 i a}]
+ e^{(k_0+6 i) a} k_0 \pi \tanh({k_0 \pi \over 4})
\right) \right) \right) \,,
\eea
where $a \equiv \arctan z$.

As an application, we now try to find the components of $v_{-,-{1\over
3}}$. This is the twist-odd eigenvector at eigenvalue $\lambda = -1/3$
whose existence is so-far unpredicted.
As we have mentioned, this state exists only when we reach 
$\lambda=-1/3$ from the right hand side. This corresponding to
$k_0\rightarrow 0$ and we find the limit
\bea
\label{G_o}
G_o(z)|_{-({1 \over 3})^+} & = & { i \sqrt{3} \over 8 \pi}[ 24
(\arctan z)^2 -\pi^2 + 6\ln(e^{-4i \arctan z}) 
\ln(1-e^{-4i \arctan z})+ 6 {\rm Li}_2[e^{-4i \arctan z}]]
\nonumber \\
& = & { 3\sqrt{3} z \over \pi}-{ 7 z^3 \over \sqrt{3} \pi}+
{43\sqrt{3} z^5 \over 25 \pi}-{ 337\sqrt{3} z^7 \over 245 \pi}
+{ 1091 z^9 \over 315 \sqrt{3} \pi}+... 
\eea
where ${\rm Li}_2[z]=\sum\limits_{k=1}^{\infty} z^k/k^2$ (for
$|z|<1$) is the {\em dilogarithm} function. 
\section{The spectrum of $M'^{12}$ and $M'^{21}$}
We digress here for a moment to present another application of our
analysis. Knowing the spectrum of $M^{'11}\equiv M'$ 
it is easy to calculate that of the matrices ${M^{\prime}}^{12}$
and ${M^{\prime}}^{21}$. 

The method is in direct
parallel to the discussions in \cite{Spec} because the matrices
${M^{\prime}}^{rs}$ obey the same useful properties as the matrices
$M^{rs}$:
\bea
&& [{M^{\prime}}^{rs}, {M^{\prime}}^{r' s'}] = 0 \label{comm}
\quad \forall r, s, r', s' = 1, 2, 3 \\
&& {M^{\prime}} + {M^{\prime}}^{12} + {M^{\prime}}^{21} =
(M^{\prime})^2 + ({M^{\prime}}^{12})^2 + ({M^{\prime}}^{21})^2 = 1
\quad , \quad {M^{\prime}}^{12} {M^{\prime}}^{21} = M^{\prime}
(M^{\prime} -1) \,.
\label{relations}
\eea

From (\ref{comm}), we see that all ${M^{\prime}}^{rs}$ share the same
eigenvectors. The continuous eigenvalues 
$\lambda^{12}(k)$ and $\lambda^{21}(k)$ of
${M^{\prime}}^{12}$ and ${M^{\prime}}^{21}$ respectively, can then be
calculated by treating (\ref{relations}) as a system of equations in
$\lambda^{rs}(k)$.  We thus obtain:
\bea
\lambda^{12}(k) - \lambda^{21}(k) &=& \pm \sqrt{(1 - \lambda(k))
(1 + 3 \lambda(k))} \\
\lambda^{12}(k) + \lambda^{21}(k) &=& 1 - \lambda(k) \,.
\eea

Because in the limit $b \rightarrow \infty$, ${M^{\prime}}^{12}$ 
and ${M^{\prime}}^{21}$ have similar diagonalized form as $M'$, 
we should obtain the same eigenvalues
as for the matrices $M^{rs}$. 
We can extend the choice of sign in front of the
square root (from \cite{Spec})\footnote{
We are using here the same definitions as in 
\cite{0105058} and \cite{Spec} for the
matrices $M^{12}$ and $M^{21}$.} to finite values of $b$. We therefore
have, for the continuous spectrum,
\bea
\lambda^{12}(k) &=& {1 \over 2} {\rm sign}(k) \sqrt{(1 - \lambda(k)) (1 +
3 \lambda(k))} +
{1 \over 2} (1 - \lambda(k)) \\
\lambda^{21}(k) &=& -{1 \over 2} {\rm sign}(k) \sqrt{(1 - \lambda(k)) (1 +
3 \lambda(k))} +
{1 \over 2} (1 - \lambda(k)) \; .
\eea
Furthermore the doubly degenerate eigenvalue $\lambda_1$ of $M'$ 
gives rise to the
following 2 eigenvalues:
\bea
\lambda_{1+}^{12} = \lambda_{1+}^{21} \equiv \lambda_{1+}  &=& {1 \over 2}
\sqrt{(1 - \lambda_1) (1 + 3 \lambda_1)} + {1 \over 2} (1 - \lambda_1) \\
\lambda_{1-}^{12} = \lambda_{1-}^{21} \equiv \lambda_{1-} &=& - {1 \over
2}
\sqrt{(1 - \lambda_1) (1 + 3 \lambda_1)} + {1 \over 2} (1 - \lambda_1) \,.
\eea
Because $\lambda_1 \in (0,1)$, we have that $\lambda_{1+} \in (0,1)$ and
$\lambda_{1-} \in \left( -{1 \over 3}, 0 \right)$.
\section{Discussions and Conclusions}
In this paper, we solved the eigenvalue and eigenvector problem for the
matrix
$M'$. We found that its spectrum is composed of a continuous spectrum,
which is
the same as the spectrum of $M$, and a new discrete spectrum, which always
contains an eigenvalue $\lambda_1$ in the range $(0, 1)$.
We obtained the closed form for all the eigenvectors and found that, for
every eigenvalue (including $-{1 \over 3}$), we have always one twist-even
state and one twist-odd state.

A particular thing that we found is that there is a critical value
$b_0=8\ln2$ above which one pair of eigenvectors in the continuous
spectrum is
replaced by one pair of eigenvectors in the discrete spectrum, although
the
eigenvalue does not change. As the parameter $b$ is claimed to be
irrelevant to the
physics\cite{0102112, Mukh}, it would be interesting to understand the
meaning of this 
critical value $b_0$.

The main difference between the spectrum of $M^{\prime}$ and that
of $M$
is that the eigenvalue $-{1 \over 3}$ is now doubly degenerate, and that
we have
one new doubly degenerate eigenvalue in the interval $(0,1)$.
As we mentioned in Section 6, 
the double degeneracy at $\lambda=-{1 \over 3}$ is a little
mysterious although we have several pieces of evidence to support it.
This degeneracy is a surprising result of
our analysis. Indeed, in the light of \cite{Moyal, Barton} it seems to
mean that we now would have {\em two} commuting coordinates in the 
Moyal product decomposition of the star product. 
It is thus worth looking closer at our
twist-odd eigenvector $v_{-,-{1\over3}}$.

Let us try to see if level truncation can help us decide if
$v_{-,-{1\over3}}$ 
really is an eigenvector. 
For this we define $w(b, L) = -3 M^{\prime} v_{-,-{1\over3}}(b)$, where
$M^{\prime}$ and $v_{-,-{1\over3}}$ are truncated to level $L$. If
$v_{-,-{1\over3}}$ is an eigenvector of $M^{\prime}$ with eigenvalue
$-{1\over3}$, we expect that $w(b, L \rightarrow \infty) =
v_{-,-{1\over3}}(b)$ for any value of $b$. We show in the following table,
the five first nonzero components of $w(b=1)$ at various levels of
truncation, as well as their values extrapolated from a fit of the
form $a_0 + a_1/\log(L) + a_2/\log(L)^2 + a_3/\log(L)^3$. In the last
lines, we show their exact values as calculated from \eref{G_o}.
\begin{center}
\vspace{6pt}
\begin{tabular}{|c||c|c|c|c|c|}  \hline
$L$ & $w(b=1)_1$ & $w(b=1)_3$ & $w(b=1)_5$ & $w(b=1)_7$ & $w(b=1)_9$
\\ \hline
100 & 0.119343 & 0.41301 & -0.446308 & 0.43575 & -0.416943 \\ \hline
150 & 0.120588 & 0.447491 & -0.487193 & 0.479292 & -0.461839 \\ \hline
200 & 0.121053 & 0.468778 & -0.512582 & 0.506505 & -0.490075 \\ \hline
300 & 0.121347 & 0.49465 & -0.543575 & 0.539889 & -0.524878 \\ \hline
400 & 0.121394 & 0.510341 & -0.562439 & 0.560288 & -0.546226 \\ \hline
$\infty$ & 0.0984355 & 0.799035 & -0.921314 & 0.962239 & -0.980627 \\
\hline \hline
exact value  & 0.119946 & 0.608215 & -0.681026 & 0.689616 & -0.682686 \\
\hline
\end{tabular}
\vspace{6pt}
\end{center}
Comparing the two last lines, we see that the result of the fit is about
20 to 40\% away from the exact value. Though discouraging, this
discrepancy
is not conclusive because the fitting function might not be a judicious
choice. Indeed note that the convergence is monotonic and very slow, and
the values of the fit are surprisingly far away from our finite level
values.

For comparison, we show in the next table
$w_+(b, L) = -3 M^{\prime} v_{+,-{1\over3}}(b)$ for $b=1$ in the level
truncation.
\begin{center}
\vspace{6pt}
\begin{tabular}{|c||c|c|c|c|c|}  \hline
$L$ & $w_+(b=1)_0$ & $w_+(b=1)_2$ & $w_+(b=1)_4$ & $w_+(b=1)_6$ &
$w_+(b=1)_8$
\\ \hline
100 & 0.874064 & -1.86078 & 1.27486 & -1.01287 & 0.855905 \\ \hline
150 & 0.903461 & -1.89268 & 1.30621 & -1.04427 & 0.887334 \\ \hline
200 & 0.920118 & -1.91092 & 1.32429 & -1.06252 & 0.905733 \\ \hline
300 & 0.938883 & -1.9316 & 1.34494 & -1.08348 & 0.926978 \\ \hline
400 & 0.949481 & -1.94335 & 1.35673 & -1.0955 & 0.939219 \\ \hline
$\infty$ & 1.02072 & -2.03708 & 1.46506 & -1.2184 & 1.07553 \\ \hline
\hline
exact value  & 1 & -2 & 1.41421 & -1.1547 & 1 \\ \hline
\end{tabular}
\vspace{6pt}
\end{center}
We see that it converges towards the expected value much better
than the $C$-odd vector does.
We can try to compare this difference in numerical behavior to the case
of the matrix $M$. Remember that in \cite{Spec}, the authors found a
candidate $C$-even eigenvector (denoted $v^+$) of eigenvalue $-{1\over3}$,
in addition to the $C$-odd eigenvector $v^-$. This candidate was however
discarded by the authors for several reasons:
\begin{itemize}
\item $v^+$ is an eigenvector of $K_1^2$ but not of $K_1$.
\item The set of eigenvectors without $v^+$ already forms a complete
basis \cite{Spec_Oku_1}.
\item The norm of $v^+$ has a worse divergence than the norm of $v^-$.
\item $v^+$ never appears in the level truncation.
\end{itemize}
Our analysis does not allow us to generalize these two first arguments to
our case\footnote{In principle, it should be possible to check the
completeness, but until now we haven't been able to simplify the algebra
involved.}. But we can do the same level truncation tests as above
with the vectors $v^+$ and $v^-$. In the following table, we show
$u^+ \equiv -3 M v^+$ at various truncations levels as well as the
expected values.
\begin{center}
\vspace{6pt}
\begin{tabular}{|c||c|c|c|c|c|}  \hline
$L$ & $(u^+)_2$ & $(u^+)_4$ & $(u^+)_6$ & $(u^+)_8$ & $(u^+)_{10}$
\\ \hline
100 & 1.12259 & -1.00375 & 0.90241 & -0.822676 & 0.758504 \\ \hline
150 & 1.1771 & -1.06413 & 0.965334 & -0.886898 & 0.823385 \\ \hline
200 & 1.21013 & -1.10103 & 1.00408 & -0.926709 & 0.863861 \\ \hline
300 & 1.24965 & -1.14547 & 1.051 & -0.975174 & 0.913374 \\ \hline
400 & 1.27328 & -1.17218 & 1.07933 & -1.00457 & 0.943521 \\ \hline
$\infty$ & 1.65594 & -1.63066 & 1.58908 & -1.55458 & 1.46357 \\ \hline
\hline
$v^+$  & 1.41421 & -1.33333 & 1.25196 & -1.18525 & 1.13039 \\ \hline
\end{tabular}
\vspace{6pt}
\end{center}
Now we compare this to the same analysis done with $u^- \equiv -3 M v^-$.
\begin{center}
\vspace{6pt}
\begin{tabular}{|c||c|c|c|c|c|}  \hline
$L$ & $(u^-)_1$ & $(u^-)_3$ & $(u^-)_5$ & $(u^-)_7$ & $(u^-)_9$
\\ \hline
100 & 0.957182 & -0.531943 & 0.399565 & -0.328769 & 0.283024 \\ \hline
150 & 0.967172 & -0.542301 & 0.410247 & -0.33963 & 0.293974 \\ \hline
200 & 0.97283 & -0.548233 & 0.416417 & -0.345949 & 0.300388 \\ \hline
300 & 0.979205 & -0.554972 & 0.423469 & -0.353213 & 0.307799 \\ \hline
400 & 0.982806 & -0.558804 & 0.4275 & -0.357384 & 0.312072 \\ \hline
$\infty$ & 1.00694 & -0.590333 & 0.46527 & -0.400556 & 0.360073 \\ \hline
\hline
$v^-$  & 1 & -0.57735 & 0.447214 & -0.377964 & 0.333333 \\ \hline
\end{tabular}
\vspace{6pt}
\end{center}
We see that the difference in numerical behavior between
$v_{-, -{1\over3}}$ and $v_{+, -{1\over3}}$ is qualitatively
similar to the difference in numerical behavior between
$v^+$ and $v^-$. This suggests that we should be suspicious
about $v_{-, -{1\over3}}$. The eigenstate indeed deserves further 
investigation. However, as we have shown in the
limit $b\rightarrow \infty$, we do believe the existence of
the state $v_{-, -{1\over3}}$. We think that the reason why 
the level truncation does not work is that the components of
$v_{-, -{1\over3}}$ do not decay fast enough and
level truncation is not very trustable in this case.

Let us move onto the other eigenvalues.
The existence of the discrete eigenvalue $\lambda_1$ in the range $(0,1)$ 
can be considered as the result of us adding  zero modes into the matrix
$M$ to get $M'$. This relationship may help us to understand the 
physical meaning of these discrete states. 
As a check, we can calculate the eigenvalues numerically in the level
truncation scheme. We found that the eigenvalue in region $(0,1)$ 
converges very fast
as the level is increased; this situation is very different from that
for $\lambda = - { 1\over 3}$ for example, which converges only
logarithmically in
level truncation \cite{0102112, Spec}. To illustrate this, we write in the
following table the value of $\lambda_1$ at $b = 0.2$, $b =1$ and $b = 5$,
found at
various levels of truncation, as well as its exact values calculated from
\eref{implicit}.
\\ \\
\begin{tabular}{|c||c|c|c|c|c||c|}  \hline
level & 1  & 5  &  10  &  50  & 100 & exact value \\  \hline
$\lambda_1(b=0.2)$  & 0.78606702 & 0.80099138 & 0.80260995 & 0.80326016
& 0.80328899 & 0.80329559 \\ \hline
$\lambda_1(b=1)$  & 0.39394374 & 0.40376417 & 0.40407525 & 0.40411239
& 0.40412026  & 0.40412740 \\ \hline
$\lambda_1(b=5)$  & 0.01082671 & 0.02795012 & 0.02859612 & 0.02873404
& 0.02873526  & 0.02873810 \\ \hline
\end{tabular}
\\ \\
We see that, at level $10$, the relative error is less than $1 \%$. And
for $b = 0.2$ and $b=1$, level $1$ is already a good approximation.

We hope that the results of this paper can find useful applications.
In particular they should lead to some information about the instantonic
sliver \cite{0105059, 0106036}. Some future work could consist of seeking
a better understanding of the density of eigenvalues in the continuous
spectrum.
Indeed, we have found no convincing argument to claim that it should be
the
same as for the matrix $M$. In fact, if those densities were the same, we
could
simplify the continuous spectrum between the numerator and the denominator
of
the ratio
$$
R = {T_p \over 2 \pi \sqrt{\alpha^{\prime}} T_{p+1}} =
{3 \left(V_{00}^{rr} + {b \over 2} \right)^2 \over \sqrt{2 \pi b^3}}
{\det (1-M^{\prime})^{3 \over 4} (1+3 M^{\prime})^{1 \over 4} \over
\det (1-M)^{3 \over 4} (1+3 M)^{1 \over 4}} \,.
$$
But this would lead to a puzzle because $M'$ has two eigenvectors with
eigenvalue
$-{1 \over 3}$ (at least we think so), whereas $M$ has only one; 
$R$ would then naively be zero
(and we know that it is one \cite{Spec_Oku_2}).

As another direction for future research we can find the spectrum of
the $M'$ matrix in the presence of a background $B$-field in the vein
of \cite{Bonora}. This will be addressed in a forthcoming work
\cite{us-future}. We can also discuss the relationship of the Moyal
product with Witten's star product in the case of including the zero
modes as in \cite{Moyal}.
\newpage
\section*{Acknowledgements}
We would like to extend our sincere gratitude to Professor Barton
Zwiebach for suggesting this project to us and for many insightful
discussions and comments. We would also like to thank Ian Ellwood for
stimulating conversation, and Professor Wati
Taylor for discussion of \cite{MT}.
We gratefully acknowledge Dmitri Belov for very useful
conversations and suggestions to the first draft.
This Research was supported in part by
the CTP and LNS of MIT and the U.S. Department of Energy 
under cooperative research agreement \# DE-FC02-94ER40818.
N.~M. and Y.-H.~H. are also supported by the Presidential 
Fellowship of MIT.

\bibliographystyle{JHEP}

\end{document}